\input phyzzx
\def\half{{1\over 2}} 
\def\dg{\dagger} 

\Pubnum={UUITP-06/97\cr 10A 97-05}
\date={8 April 1997}
\titlepage
\title{Unitary matrix model for toroidal compactifications of M theory}
\bigskip
\author {Alexios P. Polychronakos\footnote\dagger
{poly@calypso.teorfys.uu.se}}
\address{Theoretical Physics Dept., University of Ioannina
\break 45110 Ioannina, Greece}
\andaddress{Theoretical Physics Dept., Uppsala University\break
S-751 08 Uppsala, Sweden}
\bigskip
\abstract{A unitary matrix model is proposed as the large-N
matrix formulation of M theory on flat space with toroidal
topology. The model reproduces the motion of elementary D-particles
on the compact space, and admits membrane states with nonzero
wrapping around nontrivial 2-tori even at finite N. 
}

\vfill
\endpage

\def\NP{{\it Nucl. Phys.\ }}
\def\PL{{\it Phys. Lett.\ }}

\def\PRL{{\it Phys. Rev. Lett.\ }}

\def\JMP{{\it J. Math. Phys.\ }}

\def\IJMP{{\it Int. Jour. Mod. Phys.\ }}
\def\MPL{{\it Mod. Phys. Lett. A\ }}

\REF\BFSS{T.~Banks, W.~Fischler, S.H.~Shenker and L.~Susskind,
hep-th/9610043.}
\REF\CH{M.~Claudson and M.B.~Halpern, \NP {\bf B250} (1985) 689.}
\REF\BRR{M.~Baake, M.~Reinicke and V.~Rittenberg, \JMP {\bf 26} (1985) 1070.}
\REF\dHN{B.~de Witt, J.~Hoppe and H.~Nicolai, \NP {\bf B305} (1988) 545.}
\REF\DLP{J.~Dai, R.G.~Leigh and J.~Polchinski, \MPL {/bf A4} (1989) 2073.}
\REF\L{R.G.~Leigh, \MPL {\bf A4} (1989) 2767.}
\REF\P{J.~Polchinski, \PRL {bf 75} (1995) 4724; TASI96 Lectures on D-branes,
hepth/9611050.}
\REF\CB{C.~Bachas, \PL {\bf B374} (1996) 37.}
\REF\W{E.~Witten, \NP {\bf B460} (1996) 335.}
\REF\DFS{U.H.~Danielsson, G.~Ferretti and B.~Sundborg, \IJMP {/bf A11}
(1996) 5463.}
\REF\KP{D.~Kabat and P.~Pouliot, \PRL {/bf 77} (1996) 1004.}
\REF\DKPS{M.R.~Douglas, D.~Kabat, P.~Pouliot and S.H.~Shenker, hep-th/9608024.}
\REF\BD{M.~Berkooz and M.R.~Douglas, hep-th/9610236.}
\REF\S{L.~Susskind, hep-th/9611164.}
\REF\GRT{O.J.~Ganor, S.~Ramgoolam and W.~Taylor, hep-th/9611202.}
\REF\BSS{T.~Banks, N.~Seiberg and S.H.~Shenker, hep-th/9612157.}
\REF\KR{N.~Kim and S.-J.~Rey, hepth/9701139.}
\REF\BS{T.~Banks and N.~Seiberg, hep-th/9702187.}
\REF\DOS{M.R.~Douglas, H.~Ooguri and S.H.~Shenker, hep-th/9702203.}
\REF\T{W.~Taylor, hep-th/9611042.}
\REF\DVV{R.~Dijkgraaf, E.~Verlinde and H.~Verlinde, hep-th/9703030.}
\REF\H{J.~Hoppe, \IJMP {\bf A4} (1989) 5235.}
\REF\FFZ{D.~Fairlie, P.~Fletcher and C.~Zachos, \JMP {\bf 31} (1990) 1088.}
\REF\F{E.G.~Floratos, \PL {\bf B288} (1989) 335.}
\REF\D{M.R.~Douglas, hep-th/9703056.}
\REF\Ts{A.~Tseytlin, 9701125.}

The recently proposed matrix model approach to M-theory consists of
the dimensional reduction of 10-dimensional super-Yang-Mills (SYM) to 0+1
dimensions [\BFSS]. This model, which has appeared in the past in
different contexts [\CH-\dHN], has its origins in D-brane dynamics 
[\DLP-\W] and is known to describe the dynamics of D-particles
in the low-energy (nonrelativistic) limit [\DFS-\DKPS].
The remarkable conjecture made in [\BFSS] is that the full light-cone 
dynamics of the different physical objects 
within M-theory are imbedded in the large-N limit of the above matrix model.
This conjecture has survived a number of consistency checks [\BD-\DOS] in
a rapidly increasing literature.

The above matrix model applies to the case of flat uncompactified 
spacetime and would need be modified for other topologies.
For the case of toroidal compactifications of space, the model should
account for the interactions due to virtual strings winding around
the compact dimensions, as well as describe membranes wrapped around
compact submanifolds. The main proposal for doing that is to enlarge 
the matrix model to a (K+1)-dimensional SYM filed theory, 
where K is the number of compact dimensions [\BFSS,\T].
Although this certainly contains all the relevant degrees of
freedom [\S-\DVV], it most probably contains more than is actually 
needed at the
large-N limit. Furthermore, it takes away from the simplicity of the original
matrix model by adding an infinity of degrees of freedom. This is somewhat
undesirable. The hope would be that a matrix model with the correct dynamics
will contain, in the large-N limit, all the relevant degrees of freedom.
The purpose of this note is to point to a possible such model.

The proposed model consists of using {\it unitary} (rather than hermitian)
matrices for each compact dimension. This proposal is an especially
natural one in the context of D-brane dynamics. In fact, it is the
eigenvalues of Wilson loop elements winding around compact dimensions
of space and coupling to open strings which, in the dual picture,
become the coordinates of D-branes [\P]. In the uncompactified case
we recover a description in terms of (the constant mode of) the
gauge field. What we propose here is to preserve the Wilson element
itself as the dynamical object and write an appropriate action which,
in the large-N limit, recovers the full physical spectrum of the theory.

For simplicity, we can assume that
all 9 transverse dimensions are compactified with radii $R_i$. Then the 
Lagrangian would contain the terms
$$
L= \tr R\left\{  {R_I^2 \over 2 }  DU_i^\dg DU_i  + R_i^2 R_j^2
[U_i , U_j ] [U_i , U_j ]^\dg + \theta^T D \theta + R_i
\theta^T \gamma^i U_i^\dg [\theta , U_i ]\right\} 
\eqn\umat$$
as well as other possible higher-order terms. As usual, $DU_i = {\dot U}_i
-i[ A_0 , U_i ]$, and summation over all indices is
assumed. $A_0$ implements the gauge invariance constraint
$$
\sum_i R_i^2 (U_i^\dg {\dot U}_i - {\dot U}_i U_i^\dg )+ \theta \theta^T =0
\eqn\gc$$
In the uncompactified limit $R_i \to \infty$ the usual hermitian matrix 
model is recovered upon writing $U_i = \exp(i X_i /R_i )$.

The above model in the low-energy limit describes the motion of 
D-particles on the compact space. The potential is minimized when
the $U_i$ commute; their eigenvalues are phases $\exp(i \phi_i )$, and
$R_i \phi_i $ correspond to the coordinates of the D-particles on the
spatial torus. It is easy to see that the kinetic term plus the gauge
constraint for $\theta =0$ imply free motion for the $\phi_i$.

It is important that the present model can describe membranes wrapped around
any two-torus in space even at finite N. In [\BFSS] such membrane solutions
were identified, but the corresponding charges (winding numbers) were
activated only at the infinite-N limit and were zero for any finite N.
The construction is along similar lines as in [\dHN,\H-\F]. The difference is
that, now, the classical membrane coordinates are represented as $U_i (q,p) = 
\exp (i X_i (q,p) /R_i )$, where $q,p$ are coordinates on the light-cone
spatial world sheet of the membrane. $U_i$ are periodic in $q,p$ while
$X_i$ need not be. The membrane action is
$$
A=\int dqdp (U_i^\dg \partial_q U_i ~ U_j^\dg \partial_p U_j -
U_i^\dg \partial_p U_i ~ U_j^\dg \partial_q U_j )^2
\eqn\umem$$
while the wrapping number of the membrane is
$$
W=-\int dqdp (U_i^\dg \partial_q U_i U_j^\dg \partial_p U_j -
U_i^\dg \partial_p U_i U_j^\dg \partial_q U_j )
\eqn\uwr$$
Note that this would be zero if periodic $X_i$ were used. Since $U_i$ is
periodic in $p,q$, its Fourier decomposition will be
$$
U_i = \sum_{n,m} U_{i,nm} ~e^{iqn + ipm}
\eqn\ufou$$
where the coefficients $U_{i,nm}$ will be constrained by the fact that
$U_i$ is a phase. For a smooth imbedding only the lowest several Fourier
modes will be appreciable, so we can truncate the series to the lowest
$N \times N$ coefficients (or, rather, periodically repete them beyond,
which gives a discrete version of the membrane consisting of $N \times N$
points). From this, a corresponding matrix can be created
$$
U_i = \sum_{n,m} U_{i,nm} U^q V^p
\eqn\uu$$
where $U$ and $V$ are the fundamental ``quantum torus'' coordinates, that is,
$N \times N$ unitary matrices obeying $UV = \exp(i 2\pi /N) VU$. In general,
$U_i$ will not be exactly unitary and a particular ``normal ordering'' will
be required to unitarize it. (The same is true for the standard construction
in terms of hermitian matrices.) This will modify the coefficients in \uu\
by terms of orden $1/N$. Expressions \umem\ and \uwr\ for the membrane action
and wrapping number are reproduced in the large-N limit as the real and
imaginary part of the trace
$$
A+iW = \tr ( U_i U_j U_i^\dg U_j^\dg  -1)
\eqn\memat$$
This is exactly what appears as the potential term in the matrix model,
which thus reproduces the membrane action. A stretched wrapping membrane
will have $U_i (q,p) = \exp(i n_i q + i m_i p)$ and thus corresponds to
$U_i = U^{n_i} V^{m_i}$. For such matrices we get
$$
U_i U_j U_i^\dg U_j^\dg = e^{i 2\pi W/N}
\eqn\mwr$$
where $W={\vec n}_i \times {\vec n}_j = n_i m_j - n_j m_i$ is the wrapping
number. Therefore a better definition of the wrapping number would be the
$Z_N$-phase of the SU(N) matrix $U_i U_j U^\dg_i U^\dg_j$. Clearly the
wrapping is well-defined only when the matrix $U_i U_j U_i^\dg U_j^\dg$
has eigenvalues which are well-localized on the circle (their spread is,
say, less than $\pi$). Also, it is defined modulo N. Both properties are
relevant to the discretized finite-N description.  Note, further, that
we can construct many-membrane states, and thus reconstruct the full 
membrane Fock space in the large-N limit. For instance, the matrices
$$
U_i = U_{N_1}^{n_1} \oplus U_{N_2}^{n_2} ~,~~~ 
U_j = V_{N_1}^{m_1} \oplus V_{N_2}^{m_2}
\eqn\twomem$$
(where $U_N$, $V_N$ are the quantum torus matrices of dimensionality
N), represent two stretched membranes with winding numbers $n_1 m_1$ and 
$n_2 m_2$ around the $(i,j)$-torus,
and longitudinal momenta proportional to $N_1$ and $N_2$ respectively. 

The matrix model potential has stationary points when
$$
\sum_j U_i U_j U_i^\dg U_j^\dg = \sum_j U_j^\dg U_i U_j U_i^\dg
\eqn\stat$$
This has as only solutions the above many-membrane states we just
considered. Thus, as expected, stretched membranes
are local minima of the potential. The charge of a membrane wrapped along
the longitudinal direction and cycle $i$, on the other hand, would be
$$
W_i = \tr [ {\dot U}_i ( U_j U_i^\dg U_j^\dg - U_j^\dg U_i^\dg U_j ) ]
\eqn\long$$
plus fermionic terms.

A crucial property of the matrix model description of M-theory is supersymmetry.
The obvious generalization of the supersymmetry transformations for this model
would be
$$
\eqalign{
\delta U_i &= ( U_i \theta + \theta U_i ) \gamma_i \epsilon \cr
\delta \theta &= \half ( U_i^\dg {\dot U}_i + {\dot U}_i U_i^\dg
+\gamma_{ij} (U_i U_j U_i^\dg U_j^\dg +c.p. -h.c.) )\epsilon + {\tilde 
\epsilon} \cr}
\eqn\susy$$
For a static solution with $\theta =0$ the above transformation will preserve
half the SUSY when $U_i U_j U^\dg_i U^\dg_j$ is proportional to the identity, 
so that the dynamical part can be cancelled by the kinematical one [\BS]. This 
reproduces the stretched membrane solutions found earlier.

It can be checked, though, that the above is not an exact symmetry of the model.
In principle, we could start with the bosonic part of the lagrangian and
perform a ``supersymmetric completion''. This would in principle generate
higher order terms in the potential (vanishing in the uncompactified limit)
and corresponding extra terms in the SUSY transformations. We have not
attempted the supersymmetrization of the model here, so this remains an open
question. Potential problems reminiscent of fermion doubling and supersymmetry
violation on the lattice might arise here, since our matrix model is 
essentially a dimensional reduction of 10-d SYM on the dual lattice. This 
would imply that the resulting SUSY model contains infinite terms. In that 
case it may not be different than the corresponding (K+1)-dimensional SYM 
theory, with all the field modes integrated out to give an effective action 
for the `topological' degrees of freedom corresponding to Wilson loop elements
around the nontrivial spatial cycles and fermionic zero modes. Even then, 
the implication is that these remaining degrees of freedom are enough in the 
large-N limit to describe the full spectrum of M theory. 
In fact, it has been suggested in [\D] that supersymmetry may fix the
potential ordering ambiguities in the matrix model that would describe
D-brane motion on curved spaces (see also [\Ts]). If the conjecture of the
present note is correct, then, the M-theory matrix model for curved
spaces would contain unitary matrices for spaces of nontrivial connectivity,
with the form of the action fixed by supersymmetry. This, and other
consistency checks on the model, remain issues for investigation.

\ack{I would like to thank E.~Floratos, V.P.~Nair and K.~Stelle for 
discussions. I would also like to thank Imperial College and Rockefeller
University for their hospitality while this work was done.}

\refout
\end